\documentclass[12pt,aps,prd,preprint,tightenlines,superscriptaddress,showpacs,nofootinbib]
{revtex4-1}

\usepackage{epsfig}
\usepackage{amssymb,amsmath}
\usepackage{graphicx, caption}
\usepackage{epstopdf}
\usepackage{setspace}
\newcommand{\bea}{\begin{eqnarray}}
\newcommand{\eea}{\end{eqnarray}}

 \makeatletter
\def\alt{\mathrel{\mathpalette\gl@align<}}
\def\agt{\mathrel{\mathpalette\gl@align>}}
\def\gl@align#1#2{\lower.6ex\vbox{\baselineskip\z@skip\lineskip\z@
\ialign{$\m@th#1\hfil##\hfil$\crcr#2\crcr\sim\crcr}}} \makeatother

\begin{document}

\begin{center}
\baselineskip 20pt {\Large\bf
Split supersymmetry and hybrid inflation \\
in light of Atacama Cosmology Telescope DR6 data 
} 
\vspace{1cm}

{\large
Nobuchika Okada$^{~a}$ %\footnote{E-mail:okadan@ua.edu}  
 and Qaisar Shafi$^{~b}$ %\footnote{ E-mail:shafi@bartol.udel.edu}
} \vspace{.5cm}

{\baselineskip 20pt \it
$^b$Department of Physics and Astronomy,\\
University of Alabama, Tuscaloosa, AL 35487, USA \\
\vspace{2mm}
$^a$Bartol Research Institute, Department of Physics and Astronomy, \\
University of Delaware, Newark, DE 19716, USA
}
\vspace{.5cm}

\vspace{1.5cm} {\bf Abstract}
\end{center}

Inspired by the recent measurement of the scalar spectral index, $n_s = 0.9743 \pm 0.0034$,
   by the Atacama Cosmology Telescope (ACT) DR6 data, we present an update on the split supersymmetry hybrid inflation model,  
   also known as $\mu$-term hybrid inflation. 
The model employs a canonical K\"{a}hler potential but incorporates an additional renormalizable term in the superpotential $W$, 
   which yields the MSSM $\mu$-term following supersymmetry breaking. 
This additional term in $W$ is responsible for a high reheat temperature, $T_r \gtrsim 10^{12}$ GeV, 
   and consequently the necessity of split supersymmetry in this class of models. 
The predicted scalar spectral index is in excellent agreement with the ACT measurement and $r$, 
   the tensor to scalar ratio, is estimated to be less than or of order $10^{-2} -10^{-3}$. 
For the running of the scalar spectral index we find $|dn_s/d \ln k| ={\cal O}(10^{-4})$. 
With $T_r  \gtrsim 10^{12}$ GeV, leptogenesis is readily implemented in this class of models. 
A wino-like LSP with mass of around 2 TeV is a plausible dark matter candidate.

\thispagestyle{empty}

%\bigskip
\newpage

\addtocounter{page}{-1}

%%%%%%%%%%%%%%%%%%%%%%%%%%
%\baselineskip 30pt
% Main body
%%%%%%%%%%%%%%%%%%%%%%%%%%
\baselineskip 18pt
%%%%%%%%%%%%%%%%%%%%%%%%%%

%%%%%%%%%%%%%%%%%%%%%%%%%%%%%%%%%%%%%%%%%%%%%%%
%\section{Introduction}
%%%%%%%%%%%%%%%%%%%%%%%%%%%%%%%%%%%%%%%%%%%%%%%

%%%%%%%%%%%%
%\section{Introduction} 
%%%%%%%%%%%%
Supersymmetric hybrid inflation models \cite{Dvali:1994ms, Copeland:1994vg, Rehman:2009nq}, 
   based on some appropriate gauge symmetry breaking in the early universe, 
   provide an attractive framework for realizing the inflationary scenario. 
The simplest model utilizes a minimal renormalizable superpotential and a canonical K\"{a}hler potential. 
The inclusion of radiative corrections and soft supersymmetry breaking terms that appear during inflation, 
   as well as supergravity contributions, yields an inflationary potential with a prediction for the scalar spectral index 
   which is in perfectly good agreement with the recent Atacama Cosmology Telescope (ACT) measurement,  
   $n_s = 0.9743 \pm 0.0034$ \cite{ACT:2025fju, ACT:2025tim}. 
The tensor to scalar ratio $r$ in these minimal models turns out to be tiny though, namely $r \lesssim 10^{-11}$ \cite{Rehman:2025fja, Ahmad:2025mul}.

An intriguing variant of this minimal model is the so-called `$\mu$-term hybrid inflation' model \cite{Dvali:1997uq, King:1997ia}, 
   which we refer to here as `split supersymmetry hybrid inflation’ (SSHI) model \cite{Okada:2015vka}. 
Although SSHI is also based on a canonical K\"{a}hler potential, a key difference stems from the inclusion in SSHI of a renormalizable term 
   in the superpotential, which yields, after supersymmetry breaking, the $\mu$-term in the minimal supersymmetric Standard Model (MSSM). 
This additional superpotential term does not play any role during inflation, but it is important during the reheating phase 
   such that one ends up with a reheat temperature $T_r \gtrsim 10^{12}$ GeV. 
In order to avoid the gravitino problem and realize a consistent framework, one is led 
   to split supersymmetry \cite{Arkani-Hamed:2004ymt}
   with the MSSM parameter estimated to be $\mu \gtrsim 10^7$ GeV. 
As it turns out, the scalar spectral index in SSHI is in excellent agreement with the recent ACT measurement. 
Moreover, in contrast to the minimal supersymmetric hybrid inflation model, the tensor to scalar ratio $r$ in SSHI 
   turns out to be orders of magnitude larger, of order $10^{-3}-10^{-2}$ or so, and therefore testable in future experiments. 
The running of the scalar spectral index $|dn_s / d \ln k| = {\cal O}(10^{-4})$. 
Thermal leptogenesis can be readily implemented in SSHI models and a wino-like LSP with mass of around 2 TeV is a plausible dark matter candidate.

The $\mu$-term hybrid inflation is associated with the spontaneous breaking of a gauge symmetry $G$. 
Imposing an $R$ symmetry,  the renormalizable superpotantial relevant for inflation is uniquely determined as 
\begin{equation}
     W = S ( \kappa \bar{\Phi}\Phi - \kappa M^2 + \lambda H_u H_d ), 
\label{HBI_P}     
\end{equation}
where $\Phi$ and $\bar{\Phi}$ denote a conjugate pair of chiral superfields,  
   whose vacuum expectation values (VEVs) induce the spontaneous breaking of $G$ 
   down to the MSSM gauge group without supersymmetry breaking, 
   and $H_{u, d}$ is a symbolic expression of Higgs fields under the gauge group $G$ 
   which include the up-type and down-type Higgs doublets in the MSSM as their components. 
By suitable field redefinitions, we can make the dimensionless coefficients $\lambda$ and $\kappa$ to be real. 
The superpotential $W$ and the $G$-singlet superfield $S$ are assigned unit $R$-charges, 
   while the remaining superfields have zero $R$-charges. 
As we will discuss in the following, the inflation takes place along the singlet field direction, 
   and the breaking of $G$ coincides with the end of inflation. 
Therefore, the gauge symmetry $G$ should be free from the cosmological problems 
   associated with topological defects, such as monopoles, cosmic strings and domain walls.  
In this paper, we consider a left-right symmetric model based on $G=$SU(3)$_c \times$SU(2)$_L \times$SU(2)$_R \times$U(1)$_{B-L}$, 
   and introduce $\Phi$ and $\bar{\Phi}$ in the representations,  $({\bf 2}, +1)$ and $({\bf 2}, -1)$, respectively, under SU(2)$_R \times$U(1)$_{B-L}$ 
   and $H_{u,d}$ as a bi-doublet under SU(2)$_L \times$SU(2)$_R$.

In the limit of global supersymmetry, the VEV of $S$ is zero. 
However, the effect of supersymmetry breaking in supergravity induces a linear soft supersymmetry breaking term 
   in the scalar potential, $V \supset -m_G \kappa M^2 (S + S^\dagger)$, 
   which is proportional to the gravitino mass $m_G$ \cite{Rehman:2009nq}.
With the linear term, $\langle S \rangle = \sqrt{2} \, m_G/\kappa$ at the potential minimum, 
  and the last term in Eq.~(\ref{HBI_P}) yields the $\mu$-term:  
\begin{equation}
   \mu  = \gamma m_G, 
\end{equation}
where $\gamma \equiv \lambda/\kappa$. 
It was shown in \cite{Dvali:1997uq, King:1997ia} that to implement successful inflation and the desired breaking of $G$ 
  down to the MSSM gauge symmetry, the relation $\lambda > \kappa$ must be satisfied.

Taking into account radiative corrections to a canonical K\"{a}hler potential, and 
    the linear soft supersymmetry breaking term induced by supergravity effect, 
    the inflationary potential is given by
\bea
V(\phi) = m^4 \left( 1+ A \ln \left[\frac{\phi}{\phi_k} \right] \right) - 2 \sqrt{2} m_G m^2 \phi. 
 \label{Vinf}
\eea
Here, the inflaton field $\phi$ is identified with the real part of $S$ ($\phi=\sqrt{2} {\rm Re}[S]$), and 
    the renormalization scale is set equal to the inflaton value corresponding to the CMB horizon exit ($\phi_k$).  
The imaginary part of $S$ stays constant during inflation. 
For further discussion, see Refs.~\cite{Shafi:2010jr, Senoguz:2004vu, Pallis:2013dxa, Buchmuller:2014epa} for details. 
Finally, $m = \sqrt{\kappa} M$ , and the coefficient $A  \, (\ll1)$ from 1-loop quantum corrections to the canonical K\"{a}hler potential 
    is given by
\begin{equation}
%\alpha = 2 \frac{1}{16 \pi^2} (\kappa^2 N_{\Phi} + 2 \lambda^2)  =  \frac{1}{4 \pi^2} \left( \lambda^2 + \frac{N_{\Phi}}{2} \kappa^2 \right).
A =  \frac{1}{4 \pi^2} \left( \lambda^2 + \kappa^2 \right).
\end{equation}
As shown in Ref.~\cite{Dvali:1997uq, King:1997ia}, 
  the spectral index $n_s$ is predicted to be close to $0.98$ in the absence of this linear term,  
  or equivalently, in the limit of $m_G =0$. 
Note that for inflaton field values close to $M$, the leading supergravity corrections are of order $(M/m_{P})^4$, 
  where $m_P = 2.43 \times 10^{18}$ GeV is the reduced Planck mass, 
  and therefore well suppressed. 
Also, the quadratic soft supersymmetry breaking term $m_\phi^2 \,\phi^2$
  associated with the inflaton can be ignored relative to the linear term in Eq.~(\ref{Vinf}) 
  \cite{Rehman:2009nq, Shafi:2010jr, Senoguz:2004vu, Pallis:2013dxa, Buchmuller:2014epa}.

In the following inflationary analysis, we use 
\bea
V \simeq m^4, \; \;  
V^\prime = m^4 \; \frac{A}{\phi} \left( 1 - B \frac{\phi}{\phi_k} \right), \; \; 
V^{\prime \prime}  = -m^4 \; \frac{A}{\phi^2}, \; \; 
V^{\prime \prime \prime}=2 \, m^4 \; \frac{A}{\phi^3}, 
\eea
where a prime denotes a derivative with respect to $\phi$, and 
\begin{equation} \label{beta}
 B = \frac{2\sqrt{2} \; m_G \; \phi_k}{A \; m^2} 
\end{equation}
is a dimensionless parameter.  
The slow roll parameters are given by
\bea
\epsilon(\phi) &=& \frac{1}{2}  \left( \frac{V'}{V} \right)^2  \simeq \frac{1}{2}  \; \frac{A^2}{\phi^2} \left(1-B \frac{\phi}{\phi_k}\right)^2, \;\; 
\nonumber\\
\eta(\phi) &=& \frac{V''}{V} \simeq -\frac{A}{\phi^2}, \; \;  \nonumber\\
\zeta^2(\phi) &=& \frac{V' V'''}{V^2} = \frac{2 A^2 (1-B)}{\phi^4}. 
\label{slow_role}
\eea
For convenience, we have used the Planck units where the reduced Planck mass $m_P = 2.43 \times 10^{18}$ GeV is set equal to unity. 
Since $|\eta| \gg \epsilon$ for $A \ll 1$, we define the inflaton value $\phi_e$ at the end of inflation 
  by $|\eta(\phi_e)|$ = 1, so that $\phi_e \simeq \sqrt{A}$.

%%%%%%%%%%%%%%%%%%%%%%%%%%%%%%%%%%%%%%%%%%%%%%%%%%%%%
% Fig
%%%%%%%%%%%%%%%%%%%%%%%%%%%%%%%%%%%%%%%%%%%%%%%%%%%%%
\begin{figure}[t]
  \begin{center}
   \includegraphics[width=10cm]{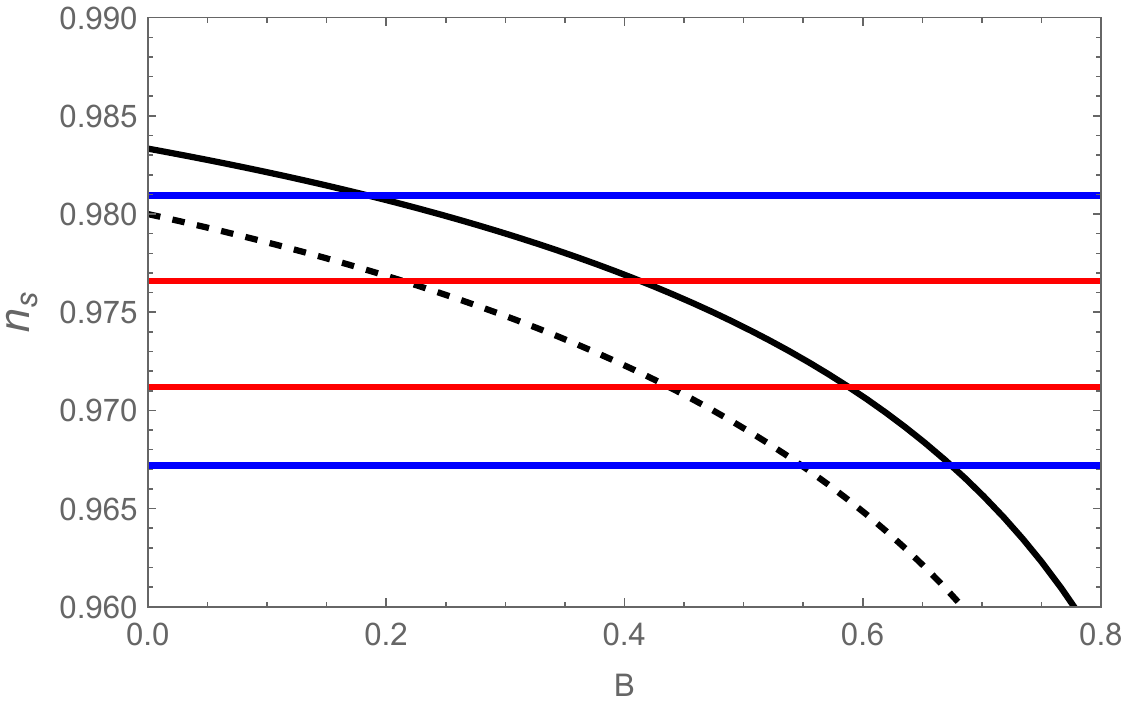}
   \end{center}
\caption{
Spectral index $n_s$ vs. $B$ for $N=60$ (solid line) and $N=50$ (dashed line). 
The region between the two red (blue) horizontal lines corresponds to 1$\sigma$ (2$\sigma$) limit 
obtained from combining the ACT DR6 with the Planck and the Baryon Acoustic Oscillation data 
\cite{ACT:2025fju, ACT:2025tim}. 
}
 \label{fig:1}
\end{figure}
%%%%%%%%%%%%%%%%%%%%%%%%%%%%%%%%%%%%%%%%%%%%%%%%%%%%%

The number of e-foldings is estimated by
\bea
N = \int_{\phi_e}^{\phi_k} \frac{V}{V'} \, d\phi
\simeq   \frac{\phi_k^2}{A} \int_{\phi_e/\phi_k}^1 \frac{x \, dx}{1-B x}
\simeq \frac{\phi_k^2}{A} \left( -\frac{1}{B} - \frac{\ln(1-B)}{B^2} \right)
\equiv \frac{\phi_k^2}{A} f(B),
\label{e-folds}
\eea
where an approximation $\phi_e/\phi_k=0$ has been used in the integral.
For the scalar spectral index, we obtain
\begin{equation}
n_s = 1 - 6\epsilon(\phi_k) + 2 \eta(\phi_k) \simeq  1 + 2 \eta(\phi_k) \simeq 1-\frac{2}{N} f(B).
\label{ns}
\end{equation}
In particular, for $N = 60$ and $B = 0$ in the limit of $m_G=0$, $f (B) = 1/2$ and hence we obtain the known result, 
  $n_s= 1 - 1/N \simeq 0.983$ in Refs.~\cite{Dvali:1997uq, King:1997ia}. 
In Figure \ref{fig:1}, we show plots of $n_s$ versus $B$ for $N=60$ (solid line) and $N=50$ (dashed line), 
  along with 1$\sigma$ (blue horizontal lines)
  and 2$\sigma$ (red horizontal lines) experimental limits after combining the ACT DR6 
  with the Planck and the Baryon Acoustic Oscillation data \cite{ACT:2025fju, ACT:2025tim}.     
To obtain the spectral index within 1$\sigma$, 
   the parameter $B$ lies in the range of $0.59 \geq B \geq 0.41$ for $N=60$, 
   while $0.44 \geq B \geq 0.22$ for $N=50$.

%%%%%%%%%%%%%%%%%%%%%%%%%%%%%%%%%%%%%%%%%%%%%%%%%%%%%
% Fig
%%%%%%%%%%%%%%%%%%%%%%%%%%%%%%%%%%%%%%%%%%%%%%%%%%%%%
\begin{figure}[h]
  \begin{center}
   \includegraphics[width=7.7cm]{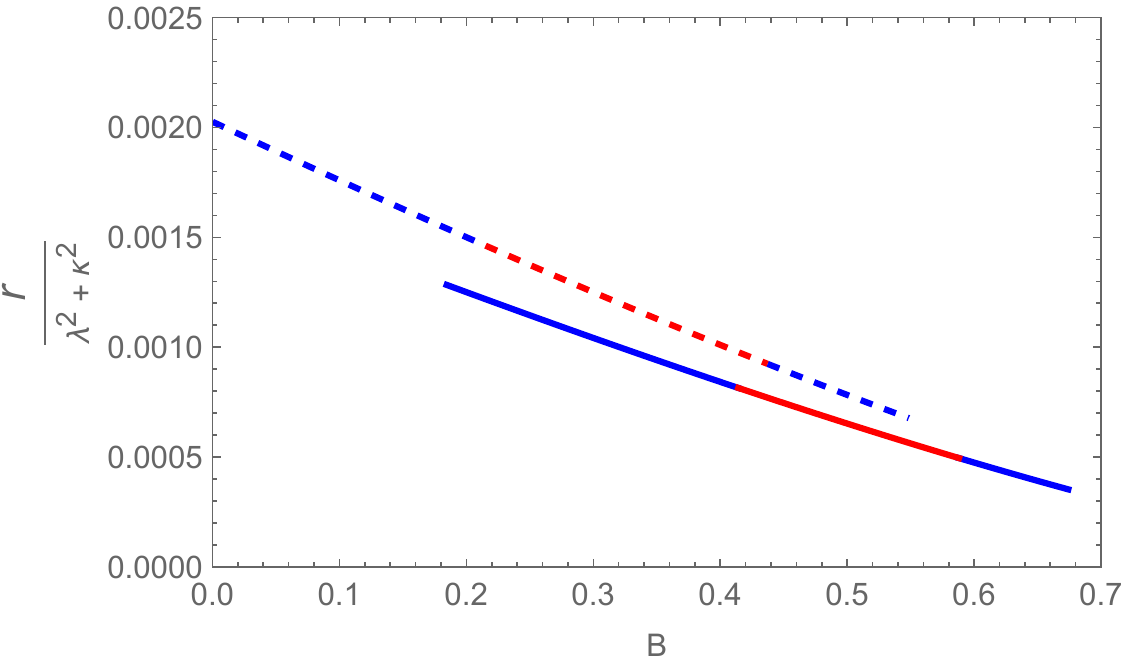} \; \;    \includegraphics[width=7.7cm]{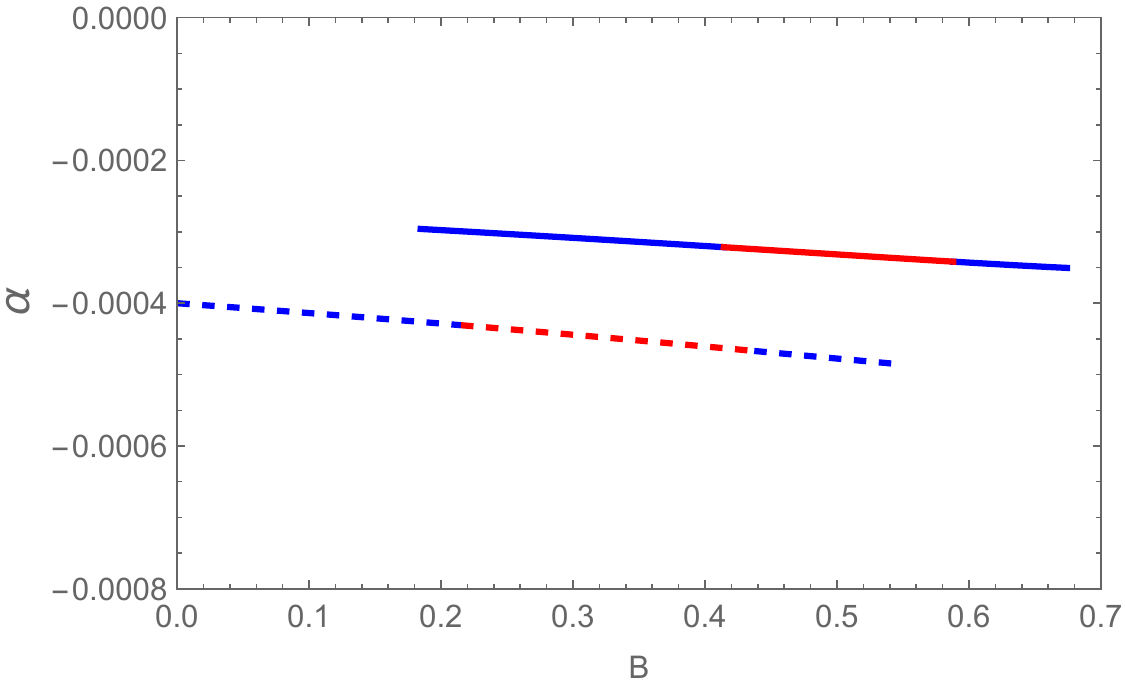}
   \end{center}
\caption{
{\bf Left Panel}: 
Tensor-to-scalar ratio normalized by $\lambda^2+\kappa^2$ as a function of $B$ for $N=60$ (solid line) and $N=50$ (dashed line).
On each line, the red (blue) region corresponds to $n_s$ within $1\sigma$ ($2 \sigma$) range shown in Figure \ref{fig:1}.  
{\bf Right Panel}: The running of the spectral index $\alpha = \frac{d n_s}{d \ln k}$ 
  as a function of $B$ for $N=60$ (solid line) and $N=50$ (dashed line).
The color coding follows that of the left panel. 
}
\label{fig:2}
\end{figure}
%%%%%%%%%%%%%%%%%%%%%%%%%%%%%%%%%%%%%%%%%%%%%%%%%%%%%
  
%%%%%%%%%%%%%%%%%%%%%%%%%%%%%%%%%%%%%%%%%%%%%%%%%%%%%
% Fig
%%%%%%%%%%%%%%%%%%%%%%%%%%%%%%%%%%%%%%%%%%%%%%%%%%%%%
\begin{figure}[t]
  \begin{center}
   \includegraphics[width=10cm]{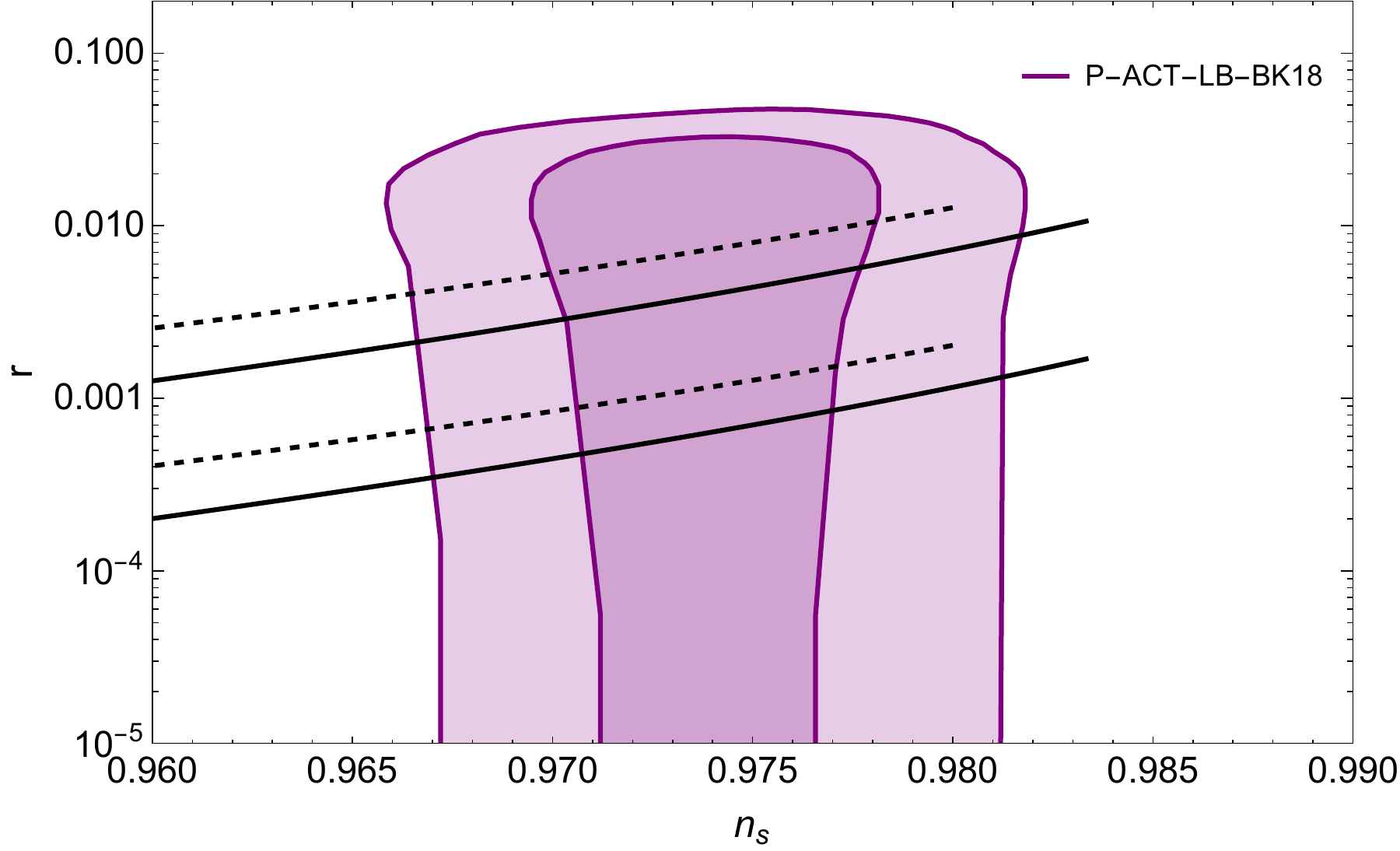}
   \end{center}
\caption{
The inflationary predictions of $\mu$-term hybrid inflation  
   along with the constraints from Planck, ACT, DESI, and BICEP/Keck data (P-ACT-LB-BK18).
The upper and lower dashed lines represent the predictions for $N=50$ 
   with the choice of $\lambda^2+\kappa^2=2 \pi$ and $1$, respectively. 
Corresponding results for $N=60$ are shown as the upper and lower solid lines. 
}
\label{fig:3}
\end{figure}
%%%%%%%%%%%%%%%%%%%%%%%%%%%%%%%%%%%%%%%%%%%%%%%%%%%%%

The tensor-to-scalar ratio is given by
\bea
r=16 \epsilon(\phi_k)   \simeq \frac{2}{\pi^2} (\lambda^2 + \kappa^2) \, (1-B)^2 \, \frac{f(B)}{N} , 
\eea
where we have used Eqs.~(\ref{slow_role}) and (\ref{e-folds}). 
In the left panel of Figure \ref{fig:2}, we show plots of $\frac{r}{\lambda^2+\kappa^2}$ versus $B$
   for $N=60$ (solid line) and $N=50$ (dashed line).
On each line, the red and blue regions correspond to $n_s$ within 1$\sigma$ and 2$\sigma$ regions 
  shown in Figure \ref{fig:1}.   
The running of the spectral index $\alpha = \frac{d n_s}{d \ln k}$ is given by
\bea
\alpha = 16 \epsilon(\phi_k) \eta(\phi_k) -24 \epsilon(\phi_k)^2  -2 \zeta^2(\phi_k) 
 \simeq 2 \zeta^2(\phi_k)  \simeq -4 \left(\frac{f(B)}{N} \right)^2 (1-B), 
\eea
where we have used $\epsilon(\phi_k)^2 \ll  \epsilon(\phi_k) |\eta(\phi_k)| \ll \zeta^2(\phi_k)$. 
The plots of $\alpha$ versus $B$ for $N=60$ (solid line) and $N=50$ (dashed line) are shown in the right panel of Figure \ref{fig:2}. 
Here, the color coding follows that of the left panel.

It is instructive to present the inflationary predictions of $\mu$-term hybrid inflation in the $(n_s, r)$-plane, 
   along with the constraints from Planck, ACT, DESI, and BICEP/Keck data (P-ACT-LB-BK18) \cite{ACT:2025tim}.  
Figure \ref{fig:3} show the inflationary predictions for $N=50$ (dashed) and $N=60$ 
  for a sample choice of $\lambda^2 +\kappa^2=1$ (upper dashed/solid line) 
  and $\lambda^2 +\kappa^2=2 \pi$ (lower dashed/solid line). 
For each line, the right end point is the prediction for $B=0$, and $n_s$ values decrease as $B$ increases 
  as shown in Figure \ref{fig:1}. 
The inner and outer shaded regions correspond to the allowed 1$\sigma$ and 2$\sigma$ regions
  from P-ACT-LB-BK18, respectively. 
The prediction with suitable values of $B$ is seen to provide an excellent fit with P-ACT-LB-BK18. 
Note that the prospective search reach of $r$ by the future CMB observations,  
   such as LiteBIRD \cite{LiteBIRD:2022cnt} and CMB-S4 \cite{Abazajian:2019eic}, 
   can approach $r$ values as low as $10^{-3}-10^{-4}$, in which case the predicted values for $\lambda^2 +\kappa^2 \gtrsim 1$ will be tested.

%%%%%%%%%%%%%%%%%%%%%%%%%%%%%%%%%%%%%%%%%%%%%%%%%%%%%
% Fig
%%%%%%%%%%%%%%%%%%%%%%%%%%%%%%%%%%%%%%%%%%%%%%%%%%%%%
\begin{figure}[h]
  \begin{center}
   \includegraphics[width=7.7cm]{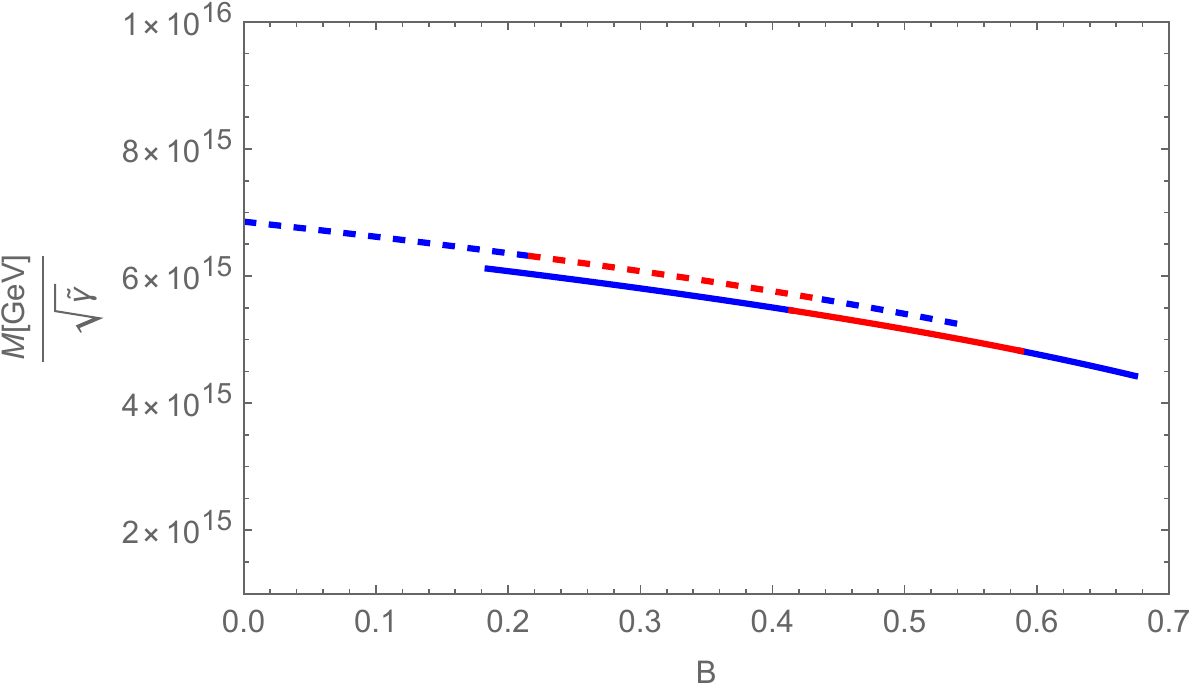} \; \;    \includegraphics[width=7.7cm]{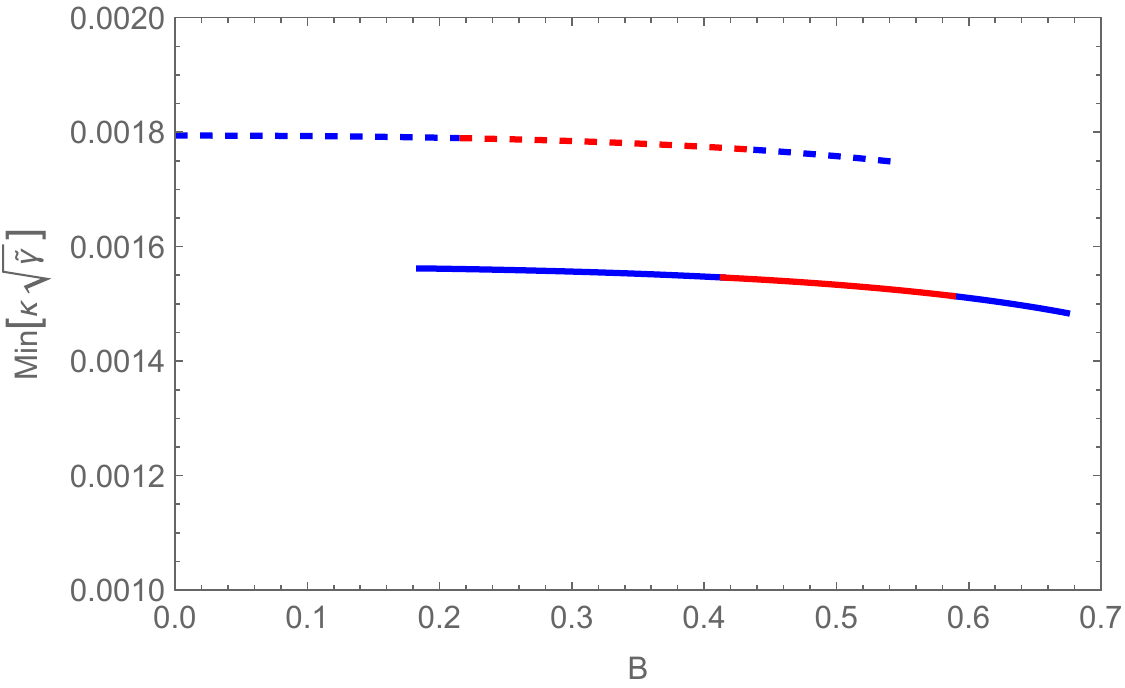}
   \end{center}
\caption{
{\bf Left Panel}: 
The gauge symmetry breaking scale (normalized by $\sqrt{\tilde{\gamma}}$) as a function of $B$ 
  for $N=60$ (solid line) and $N=50$ (dashed line).
The color coding follows that of Figure \ref{fig:2}. 
{\bf Right Panel}: 
The minimum value of $\kappa \sqrt{\tilde{\gamma}}$ from the theoretical requirement $\phi_k/M >1$ 
   for $N=60$ (solid line) and $N=50$ (dashed line), with the same color coding as in the left panel. 
}
\label{fig:4}
\end{figure}
%%%%%%%%%%%%%%%%%%%%%%%%%%%%%%%%%%%%%%%%%%%%%%%%%%%%%

The amplitude of the curvature perturbation $\Delta_{\cal{R}}$ is given by
\begin{equation}
  \Delta_{\mathcal{R}}^2 = \frac{1}{24 \pi^2} \frac{V(\phi_k)}{\epsilon(\phi_k)} 
  \simeq \frac{20 \left(\frac{N}{60}\right) M^4}{ \tilde{\gamma}^2(1-B)^2 f(B)} ,
\end{equation}
where $\tilde{\gamma} =  \sqrt{\gamma^2 + 1} > \sqrt{2}$ under the condition $\gamma > 1$.  
This should be compared with the measurement by 
  Planck 2018 TT, TE, EE + lowE + lensing \cite{Planck:2018jri}
  with the pivot scale $k=0.05$/Mpc, namely, $\Delta_{\mathcal{R}}^2 =2.1 \times 10^{-9}$.  
This condition leads to the following expression for the gauge symmetry breaking scale $M$:
\begin{equation}
M [{\rm GeV}] \simeq 7.8 \times 10^{15} \, \sqrt{\tilde{\gamma}}  \left(\frac{60}{N}\right)^{1/4}  (1-B)^{1/2} f(B)^{1/4}. 
\label{M_formula}
\end{equation}  
In the left panel of Figure \ref{fig:4}, we show $M/\sqrt{\tilde{\gamma}}$ 
   as a function of $B$ for $N=60$ (solid line) and $N=50$ (dashed line). 
The color coding follows that of Figure \ref{fig:2}. 
Since $\tilde{\gamma} >\sqrt{2}$, the gauge symmetry breaking scale is found to be 
  around a typical grand unification scale of $10^{16}$ GeV.

The inflaton trajectory must be bounded along the $\Phi$, $\bar{\Phi}$ directions, which requires $\phi_k > M$. 
Using Eqs.~(\ref{e-folds}) and (\ref{M_formula}), the ratio $\phi_k/M$ is given by
\begin{equation}
\phi_k / M \simeq 380 \times  \kappa \sqrt{\tilde{\gamma}} \left( \frac{N}{60} \right)^{3/4} (1-B)^{-1/2} f(B)^{-3/4}, 
\end{equation}
  and requiring it to be greater than unity yields a lower bound on $\kappa \sqrt{\tilde{\gamma}}$, 
  which is shown in the right panel of Figure \ref{fig:4}. 
The solid and dashed lines depict the lower bound for $N=60$ and $N=50$, with the same color coding as in Figure \ref{fig:2}. 
We can see that this bound weakly depends on $B$ and is roughly $\sqrt{\lambda \kappa} \gtrsim 10^{-3}$.

Next let us consider the process of reheating after inflation. 
Through the inflaton $\phi$ coupling to $H_{u, d}$ in the superpotential of Eq.~(\ref{HBI_P}), 
   the main decay mode of inflaton is to a pair of Higgsinos. 
The decay width of this process is calculated to be 
\bea
\Gamma (\phi \to  \tilde{H_u} \tilde{H_d}) = \frac{\lambda^2}{8 \pi} m_{\phi},
\eea
where $m_\phi = \sqrt{2} \kappa M$ is the inflaton mass.
In estimating the reheating temperature after inflation $T_r$, we adopt the instantaneous decay approximation, 
  $\Gamma = H(T_r) =  T_r^2\sqrt{\frac{\pi^2}{90} g_{\star}} $ with $g_{\star} = 228.75$ for the MSSM, 
  which yields 
\bea
T_{r} [{\rm GeV}] \simeq  1.5 \times 10^{16}  \,  ( \kappa \sqrt{\tilde{\gamma}} )^{3/2} \frac{\gamma}{\sqrt{\tilde{\gamma}}} 
  \left( \frac{60}{N} \right)^{1/8} (1-B)^{1/4} f(B)^{1/8} \, \gtrsim  \, 10^{12} . 
\eea
In the last expression above, we have used the lower bound on $\kappa \sqrt{\tilde{\gamma}}$ 
  shown in the right panel of Figure \ref{fig:4} and $\frac{\gamma}{\sqrt{\tilde{\gamma}}} \simeq \sqrt{\gamma} > 1$.

The cosmology of supersymmetric models requires careful consideration due to the presence of the gravitino, 
   the superpartner of the graviton, which is long-lived since it couples very weakly to other particles. 
We first consider the case in which the gravitino is the LSP and a dark matter candidate. 
The relic abundance of thermally produced gravitinos is estimated as \cite{Bolz:2000fu}:
\bea
\Omega_{G}h^2 \simeq 0.12 \left( \frac{T_r}{10^{12}\text{ GeV}} \right)
\left( \frac{150 \text{ TeV}}{m_G} \right) \left( \frac{M_3}{3 \text{ TeV}} \right)^2,
\label{Omega}
\eea
   where $M_3$ is the gluino mass. 
Given the lower bound on the reheating temperature $T_r \gtrsim 10^{12}$ GeV in our inflationary scenario, 
  and taking into account the current experimental lower bounds on the gluino mass 
  $M_3 > 1- 3$ TeV (see, for example, Ref.~\cite{ATLAS:2024lda}), 
  $m_G \gg M_3$ in Eq.~(\ref{Omega}) is required 
  to match the observed dark matter relic abundance $\Omega_{DM} h^2 =0.12$ \cite{Planck:2018vyg}. 
This contradicts the assumption that the gravitino is the LSP, and therefore, we exclude the possibility of gravitino LSP..

The lifetime of unstable gravitino is estimated to be
\bea \label{taug}
 \tau_G \simeq 1 \; {\rm sec} \times \left(  \frac{21.5\; {\rm TeV}}{m_G}\right)^3.
\eea
Since the age of the universe at the beginning of  the Big Bang Nucleosynthesis (BBN) is around a second, 
  the gravitino decays after BBN if $m_G < 21.5$ GeV. 
In this case, we encounter the cosmological gravitino problem \cite{Khlopov:1984pf, Ellis:1984eq}, 
  namely, the energetic particles created by the gravitino decay can destroy the light nuclei
  successfully synthesized during BBN.
To avoid this problem, the reheating temperature after inflation has an upper bound,
  $T_r < 10^6-10^9$ GeV for $100$ GeV$\lesssim m_G \lesssim10$ TeV.
The reheating temperature in our scenario exceeds this upper bound.

For $m_G > 21.5$ TeV,  the gravitino decays before BBN, and thus the BBN bound on the reheating temperature does not apply.
However, in this case, another cosmological constraint must be considered. 
Given the high reheating temperature in our scenario, it is natural to assume that the LSP neutralino constitutes thermal dark matter, 
   with its relic abundance determined by freeze-out from the thermal plasma. 
A long-lived gravitino decays into lighter sparticles at late times, which subsequently undergo cascade decays, producing additional LSP neutralinos.
This non-thermal contribution can potentially lead to an overproduction of neutralino dark matter, exceeding the observed dark matter abundance.
A simple way to avoid this issue is to require that the gravitino decays before the LSP neutralino freezes out from the thermal plasma.
In such a case, the neutralinos produced by gravitino decay thermalize immediately, and the final LSP abundance becomes 
  independent of the early gravitino population.
Using a typical value of the ratio $x_F\equiv m_{{\tilde \chi}^0}/T_d \simeq 20$,
  where $T_d$ is the freeze-out temperature of the LSP neutralino with mss $m_{{\tilde \chi}^0}$, 
   this condition corresponds to a gravitino lifetime of
\bea
  \tau_G \lesssim  10^{-10} \; {\rm sec} \times  
   \left( \frac{2 \text{ TeV}}{m_{{\tilde \chi}^0}} \right)^2.
\eea
Combining this with Eq.(\ref{taug}), we find the lower bound on the gravitino mass: 
\bea
  m_G \gtrsim 4.6 \times 10^7 \text{ GeV} \left( \frac{m_{{\tilde \chi}^0}}{2 \text{ TeV}} \right)^{2/3}.
\label{LB_mG}
\eea
Therefore, our cosmological scenario favors a gravitino mass at an intermediate scale.

In our mechanism for generating the $\mu$-term, its scale is set by the gravitino mass,
   such that $\mu = \gamma m_G$ with $\gamma \gtrsim 1$.
To achieve electroweak symmetry breaking, the soft supersymmetry breaking mass for the MSSM Higgs doublets 
    must satisfy the condition $|m_0^2| \gtrsim \mu^2$. 
This implies that the soft scalar mass parameter $m_0$ is also of intermediate scale.
Meanwhile, the mass scale of the dark matter LSP neutralino must lie in the range of 100 GeV to a few TeV 
   to account for the observed relic abundance. 
This setup is naturally compatible with the so-called split supersymmetry scenario \cite{Arkani-Hamed:2004ymt}, 
  which features a hierarchy between the scalar superpartner masses and the gaugino masses.
In particular, in the split supersymmetry scenario with a large $\mu$-term, 
  a wino-like LSP with a mass around 2 TeV emerges as the simplest dark matter candidate \cite{Giudice:2004tc, Arkani-Hamed:2004zhs}.

%%%%%%%%%%%%%%%%%%%%%%%%%%%%%%%%%%%%%%%%%%%%%%%%%%%%%
% Fig
%%%%%%%%%%%%%%%%%%%%%%%%%%%%%%%%%%%%%%%%%%%%%%%%%%%%%
\begin{figure}[h]
  \begin{center}
   \includegraphics[width=10 cm]{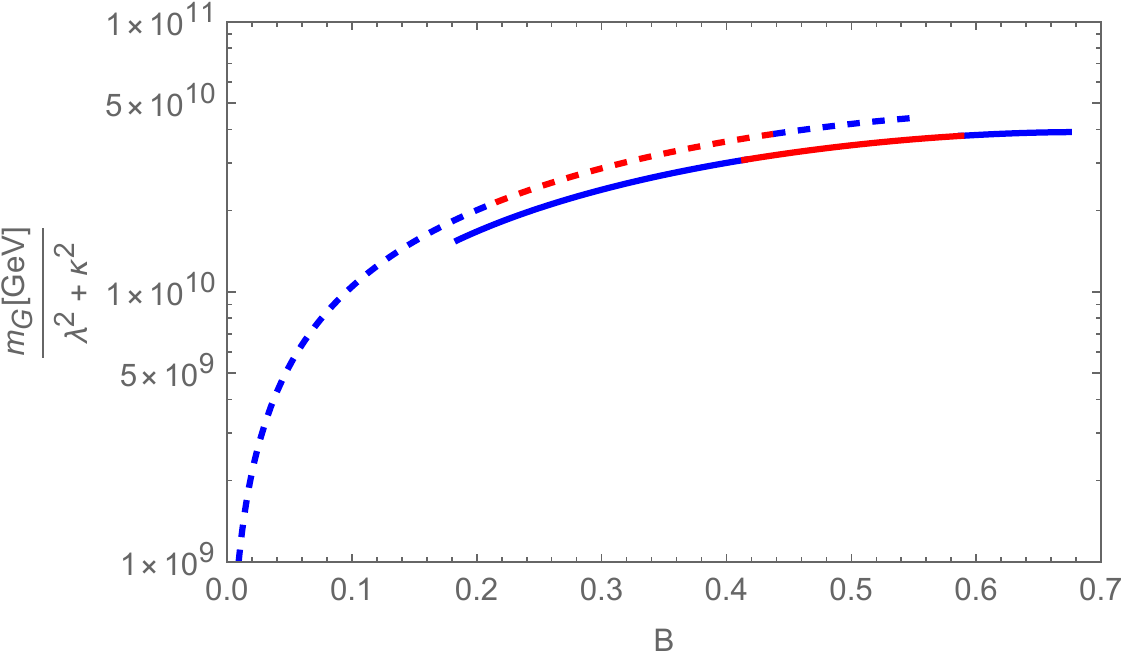}
   \end{center}
\caption{
Gravitino mass normalized by $\lambda^2+\kappa^2$ as a function of $B$ for $N=60$ (solid line) and $N=50$ (dashed line).
The color coding follows that of Figure \ref{fig:2}. 
}
\label{fig:5}
\end{figure}
%%%%%%%%%%%%%%%%%%%%%%%%%%%%%%%%%%%%%%%%%%%%%%%%%%%%%

By substituting $\phi_k$ from Eq.~(\ref{e-folds}) and $M$ from Eq.~(\ref{M_formula}) into the definition of $B$ 
 in Eq.~(\ref{beta}), we can express the gravitino mass in a simple form: 
\bea
   m_G[{\rm GeV]} = 1.8 \times 10^{11} \; (\lambda^2 + \kappa^2) \, B  \,(1-B) \, f(B)  \,\left( \frac{60}{N} \right). 
\eea
In Figure \ref{fig:5}, we show the gravitino mass (normalized by $\lambda^2 + \kappa^2$) as a function of $B$ 
  for $N=60$ (solid line) and $N=50$ (dashed line). 
The color coding follows that of Figure \ref{fig:2}. 
The lower bound on the gravitino mass in Eq.~(\ref{LB_mG}) requires $\lambda^2 + \kappa^2 \gtrsim 10^{-3}$ 
  for the 1$\sigma$ allowed regions in red.  
Since $m_G \propto B$, the condition of Eq.~(\ref{LB_mG}) excludes a $B$ value very close to 0 for $N=50$.

Finally, by introducing higher-dimensional operators into the superpotential, such as
\bea
W \supset \frac{L^c L^c  \bar{\Phi} \bar{\Phi}}{m_P},
\eea
where $L^c$ is an SU(2)$_R$ anti-lepton doublet with U(1)$_{B-L}$ charge $+1$,
   Majorana mass terms for the right-handed neutrinos are generated
   in association with the breaking of SU(2)$_R \times$ U(1)$_{B-L}$.
As usual, after electroweak symmetry breaking,
   the tiny Majorana masses for the light neutrinos are generated via the type-I seesaw mechanism 
   \cite{Minkowski:1977sc, Yanagida:1979as, GellMann:1980vs, Mohapatra:1979ia}.
In our scenario, the reheating temperature after inflation, $T_r \gtrsim 10^{12}$ GeV,
   is sufficiently high to support successful thermal leptogenesis~\cite{Fukugita:1986hr}.

In conclusion, starting with a minimal supersymmetric hybrid inflation model, the inclusion of an additional renormalizable term 
   in the superpotential yields the MSSM $\mu$-term following supersymmetry breaking. 
 The implications of this seemingly straightforward extension of the minimal model are far reaching 
    for supersymmetry and the inflationary scenario.  
A high reheat temperature coupled with the gravitino constraint requires split supersymmetry for consistency. 
The scalar spectral index $n_s$ in this ``$\mu$-term hybrid inflation" turns out to be in excellent agreement 
   with the ACT measurement and $r$, the tensor to scalar ratio, can approach values as high as $10^{-3}-10^{-2}$. 
The running of the scalar spectral index is estimated to be
   $|dn_s/d lnk| ={\cal O}(10^{-4})$. 
A wino-like LSP with mass of around 2 TeV or so is a plausible dark matter candidate.

%%%%%%%%%%%%%%%%%%%%%%%%%%%%%%%%%%
\section*{Acknowledgments}
%%%%%%%%%%%%%%%%%%%%%%%%%%%%%%%%%%
The work of N.O. is supported in part by the United States Department of Energy Grants
DE-SC0012447 and DE-SC0023713.

\newpage
%%%%%%%%%%%%%%%%%%%%%%%%%%%  
\bibliographystyle{utphysII}
\bibliography{References}

\providecommand{\href}[2]{#2}\begingroup\raggedright\begin{thebibliography}{10}

\bibitem{Dvali:1994ms}
G.~R. Dvali, Q.~Shafi, and R.~K. Schaefer, ``{Large scale structure and
  supersymmetric inflation without fine tuning},''
  \href{http://dx.doi.org/10.1103/PhysRevLett.73.1886}{Phys. Rev. Lett. {\bf
  73} (1994)  1886--1889}, \href{http://arxiv.org/abs/hep-ph/9406319}{{\tt
  arXiv:hep-ph/9406319}}.

\bibitem{Copeland:1994vg}
E.~J. Copeland, A.~R. Liddle, D.~H. Lyth, E.~D. Stewart, and D.~Wands, ``{False
  vacuum inflation with Einstein gravity},''
  \href{http://dx.doi.org/10.1103/PhysRevD.49.6410}{Phys. Rev. D {\bf 49}
  (1994)  6410--6433}, \href{http://arxiv.org/abs/astro-ph/9401011}{{\tt
  arXiv:astro-ph/9401011}}.

\bibitem{Rehman:2009nq}
M.~U. Rehman, Q.~Shafi, and J.~R. Wickman, ``{Supersymmetric Hybrid Inflation
  Redux},'' \href{http://dx.doi.org/10.1016/j.physletb.2009.12.010}{Phys. Lett.
  B {\bf 683} (2010)  191--195}, \href{http://arxiv.org/abs/0908.3896}{{\tt
  arXiv:0908.3896 [hep-ph]}}.

\bibitem{ACT:2025fju}
{\bf ACT} Collaboration, T.~Louis {\em et al.}, ``{The Atacama Cosmology
  Telescope: DR6 Power Spectra, Likelihoods and $\Lambda$CDM Parameters},''
  \href{http://arxiv.org/abs/2503.14452}{{\tt arXiv:2503.14452 [astro-ph.CO]}}.

\bibitem{ACT:2025tim}
{\bf ACT} Collaboration, E.~Calabrese {\em et al.}, ``{The Atacama Cosmology
  Telescope: DR6 Constraints on Extended Cosmological Models},''
  \href{http://arxiv.org/abs/2503.14454}{{\tt arXiv:2503.14454 [astro-ph.CO]}}.

\bibitem{Rehman:2025fja}
M.~U. Rehman and Q.~Shafi, ``{Supersymmetric hybrid inflation in light of the
  Atacama Cosmology Telescope data release 6, Planck 2018, and LB-BK18},''
  \href{http://dx.doi.org/10.1103/mwn8-rnsx}{Phys. Rev. D {\bf 112} (2025)
  no.~2, 023529}, \href{http://arxiv.org/abs/2504.14831}{{\tt arXiv:2504.14831
  [astro-ph.CO]}}.

\bibitem{Ahmad:2025mul}
M.~N. Ahmad and M.~U. Rehman, ``{Supersymmetric Hybrid Inflation with
  K{\"a}hler-Induced $\mathbf{R}$-Symmetry Breaking},''
  \href{http://arxiv.org/abs/2506.23244}{{\tt arXiv:2506.23244 [hep-ph]}}.

\bibitem{Dvali:1997uq}
G.~R. Dvali, G.~Lazarides, and Q.~Shafi, ``{Mu problem and hybrid inflation in
  supersymmetric SU(2)-L x SU(2)-R x U(1)-(B-L)},''
  \href{http://dx.doi.org/10.1016/S0370-2693(98)00145-2}{Phys. Lett. B {\bf
  424} (1998)  259--264}, \href{http://arxiv.org/abs/hep-ph/9710314}{{\tt
  arXiv:hep-ph/9710314}}.

\bibitem{King:1997ia}
S.~F. King and Q.~Shafi, ``{Minimal supersymmetric SU(4) x SU(2)-L x
  SU(2)-R},'' \href{http://dx.doi.org/10.1016/S0370-2693(98)00058-6}{Phys.
  Lett. B {\bf 422} (1998)  135--140},
  \href{http://arxiv.org/abs/hep-ph/9711288}{{\tt arXiv:hep-ph/9711288}}.

\bibitem{Okada:2015vka}
N.~Okada and Q.~Shafi, ``{$\mu$-term hybrid inflation and split
  supersymmetry},''
  \href{http://dx.doi.org/10.1016/j.physletb.2017.11.015}{Phys. Lett. B {\bf
  775} (2017)  348--351}, \href{http://arxiv.org/abs/1506.01410}{{\tt
  arXiv:1506.01410 [hep-ph]}}.

\bibitem{Arkani-Hamed:2004ymt}
N.~Arkani-Hamed and S.~Dimopoulos, ``{Supersymmetric unification without low
  energy supersymmetry and signatures for fine-tuning at the LHC},''
  \href{http://dx.doi.org/10.1088/1126-6708/2005/06/073}{JHEP {\bf 06} (2005)
  073}, \href{http://arxiv.org/abs/hep-th/0405159}{{\tt arXiv:hep-th/0405159}}.

\bibitem{Shafi:2010jr}
Q.~Shafi and J.~R. Wickman, ``{Observable Gravity Waves From Supersymmetric
  Hybrid Inflation},''
  \href{http://dx.doi.org/10.1016/j.physletb.2011.01.002}{Phys. Lett. B {\bf
  696} (2011)  438--446}, \href{http://arxiv.org/abs/1009.5340}{{\tt
  arXiv:1009.5340 [hep-ph]}}.

\bibitem{Senoguz:2004vu}
V.~N. Senoguz and Q.~Shafi, ``{Reheat temperature in supersymmetric hybrid
  inflation models},''
  \href{http://dx.doi.org/10.1103/PhysRevD.71.043514}{Phys. Rev. D {\bf 71}
  (2005)  043514}, \href{http://arxiv.org/abs/hep-ph/0412102}{{\tt
  arXiv:hep-ph/0412102}}.

\bibitem{Pallis:2013dxa}
C.~Pallis and Q.~Shafi, ``{Update on Minimal Supersymmetric Hybrid Inflation in
  Light of PLANCK},''
  \href{http://dx.doi.org/10.1016/j.physletb.2013.07.029}{Phys. Lett. B {\bf
  725} (2013)  327--333}, \href{http://arxiv.org/abs/1304.5202}{{\tt
  arXiv:1304.5202 [hep-ph]}}.

\bibitem{Buchmuller:2014epa}
W.~Buchm{\"u}ller, V.~Domcke, K.~Kamada, and K.~Schmitz, ``{Hybrid Inflation in
  the Complex Plane},''
  \href{http://dx.doi.org/10.1088/1475-7516/2014/07/054}{JCAP {\bf 07} (2014)
  054}, \href{http://arxiv.org/abs/1404.1832}{{\tt arXiv:1404.1832 [hep-ph]}}.

\bibitem{LiteBIRD:2022cnt}
{\bf LiteBIRD} Collaboration, E.~Allys {\em et al.}, ``{Probing Cosmic
  Inflation with the LiteBIRD Cosmic Microwave Background Polarization
  Survey},'' \href{http://dx.doi.org/10.1093/ptep/ptac150}{PTEP {\bf 2023}
  (2023) no.~4, 042F01}, \href{http://arxiv.org/abs/2202.02773}{{\tt
  arXiv:2202.02773 [astro-ph.IM]}}.

\bibitem{Abazajian:2019eic}
K.~Abazajian {\em et al.}, ``{CMB-S4 Science Case, Reference Design, and
  Project Plan},'' \href{http://arxiv.org/abs/1907.04473}{{\tt arXiv:1907.04473
  [astro-ph.IM]}}.

\bibitem{Planck:2018jri}
{\bf Planck} Collaboration, Y.~Akrami {\em et al.}, ``{Planck 2018 results. X.
  Constraints on inflation},''
  \href{http://dx.doi.org/10.1051/0004-6361/201833887}{Astron. Astrophys. {\bf
  641} (2020)  A10}, \href{http://arxiv.org/abs/1807.06211}{{\tt
  arXiv:1807.06211 [astro-ph.CO]}}.

\bibitem{Bolz:2000fu}
M.~Bolz, A.~Brandenburg, and W.~Buchmuller, ``{Thermal production of
  gravitinos},'' \href{http://dx.doi.org/10.1016/S0550-3213(01)00132-8}{Nucl.
  Phys. B {\bf 606} (2001)  518--544},
  \href{http://arxiv.org/abs/hep-ph/0012052}{{\tt arXiv:hep-ph/0012052}}.
  [Erratum: Nucl.Phys.B 790, 336--337 (2008)].

\bibitem{ATLAS:2024lda}
{\bf ATLAS} Collaboration, G.~Aad {\em et al.}, ``{The quest to discover
  supersymmetry at the ATLAS experiment},''
  \href{http://dx.doi.org/10.1016/j.physrep.2024.09.010}{Phys. Rept. {\bf 1116}
  (2025)  261--300}, \href{http://arxiv.org/abs/2403.02455}{{\tt
  arXiv:2403.02455 [hep-ex]}}.

\bibitem{Planck:2018vyg}
{\bf Planck} Collaboration, N.~Aghanim {\em et al.}, ``{Planck 2018 results.
  VI. Cosmological parameters},''
  \href{http://dx.doi.org/10.1051/0004-6361/201833910}{Astron. Astrophys. {\bf
  641} (2020)  A6}, \href{http://arxiv.org/abs/1807.06209}{{\tt
  arXiv:1807.06209 [astro-ph.CO]}}. [Erratum: Astron.Astrophys. 652, C4
  (2021)].

\bibitem{Khlopov:1984pf}
M.~Y. Khlopov and A.~D. Linde, ``{Is It Easy to Save the Gravitino?},''
  \href{http://dx.doi.org/10.1016/0370-2693(84)91656-3}{Phys. Lett. B {\bf 138}
  (1984)  265--268}.

\bibitem{Ellis:1984eq}
J.~R. Ellis, J.~E. Kim, and D.~V. Nanopoulos, ``{Cosmological Gravitino
  Regeneration and Decay},''
  \href{http://dx.doi.org/10.1016/0370-2693(84)90334-4}{Phys. Lett. B {\bf 145}
  (1984)  181--186}.

\bibitem{Giudice:2004tc}
G.~F. Giudice and A.~Romanino, ``{Split supersymmetry},''
  \href{http://dx.doi.org/10.1016/j.nuclphysb.2004.08.001}{Nucl. Phys. B {\bf
  699} (2004)  65--89}, \href{http://arxiv.org/abs/hep-ph/0406088}{{\tt
  arXiv:hep-ph/0406088}}. [Erratum: Nucl.Phys.B 706, 487--487 (2005)].

\bibitem{Arkani-Hamed:2004zhs}
N.~Arkani-Hamed, S.~Dimopoulos, G.~F. Giudice, and A.~Romanino, ``{Aspects of
  split supersymmetry},''
  \href{http://dx.doi.org/10.1016/j.nuclphysb.2004.12.026}{Nucl. Phys. B {\bf
  709} (2005)  3--46}, \href{http://arxiv.org/abs/hep-ph/0409232}{{\tt
  arXiv:hep-ph/0409232}}.

\bibitem{Minkowski:1977sc}
P.~Minkowski, ``{$\mu \to e\gamma$ at a Rate of One Out of $10^{9}$ Muon
  Decays?},'' \href{http://dx.doi.org/10.1016/0370-2693(77)90435-X}{Phys. Lett.
  B {\bf 67} (1977)  421--428}.

\bibitem{Yanagida:1979as}
T.~Yanagida, ``{Horizontal gauge symmetry and masses of neutrinos},'' Conf.
  Proc. C {\bf 7902131} (1979)  95--99.

\bibitem{GellMann:1980vs}
M.~Gell-Mann, P.~Ramond, and R.~Slansky, ``{Complex Spinors and Unified
  Theories},'' Conf. Proc. C {\bf 790927} (1979)  315--321,
  \href{http://arxiv.org/abs/1306.4669}{{\tt arXiv:1306.4669 [hep-th]}}.

\bibitem{Mohapatra:1979ia}
R.~N. Mohapatra and G.~Senjanovic, ``{Neutrino Mass and Spontaneous Parity
  Nonconservation},'' \href{http://dx.doi.org/10.1103/PhysRevLett.44.912}{Phys.
  Rev. Lett. {\bf 44} (1980)  912}.

\bibitem{Fukugita:1986hr}
M.~Fukugita and T.~Yanagida, ``{Baryogenesis Without Grand Unification},''
  \href{http://dx.doi.org/10.1016/0370-2693(86)91126-3}{Phys. Lett. B {\bf 174}
  (1986)  45--47}.

\end{thebibliography}\endgroup

\end{document}